\documentclass[12pt]{article}
\usepackage[T1]{fontenc}
\usepackage[latin1]{inputenc}
\usepackage{graphics}
\usepackage{verbatim}
\usepackage{cite}
\begin{document}  
\newcommand{\bd}[1]{ \mbox{\boldmath $#1$}  }
\newcommand{\xslash}[1]{\overlay{#1}{/}}
\newcommand{\sla}[1]{\xslash{#1}}
\begin{titlepage}  
\thispagestyle{empty}  
  
\begin{center}  
  
{\Large \bf Uncorrelated scattering approximation for the scattering and 
break-up of weakly bound nuclei on heavy targets}

\vspace{0.5cm} 
{\large A.M. Moro, J.A. Caballero and J. G\'omez-Camacho}

\vspace{.3cm} 
{Departamento de F\'{\i}sica At\'omica, Molecular y Nuclear, 
Universidad de Sevilla,  
Apdo. 1065, E-41080 Sevilla, Spain}\\ 

\end{center} 
 
\vspace{0.5cm} 

\begin{abstract}
The scattering of a weakly bound (halo) projectile nucleus by a heavy target 
nucleus is investigated. 
A new approach, called the Uncorrelated
Scattering Approximation, is proposed. The main approximation
involved is to neglect the 
correlation between the fragments of the projectile in the  region where the 
interaction with the target is important. The formalism makes use of 
hyper-spherical harmonics,  Raynal-Revay coefficients and momentum-localized
wave functions to expand projectile channel wave functions 
in terms of products 
of the channel wave function of the individual fragments. 
Within this approach, the kinetic energy and angular momentum of each
fragment is conserved during the scattering process. 
The elastic, inelastic and break-up S-matrices are obtained
as an analytic combination involving the bound wave function of the projectile
and the product of the S-matrices of the fragments. The approach is 
applied to describe the scattering of deuteron on $^{58}$Ni at several 
energies. The results are compared with experimental data and 
continuum-discretized coupled-channels calculations.

\vspace{0.5cm}



\end{abstract}

\end{titlepage}

\newpage

\setcounter{page}{1} 

\section{Introduction\label{section:intro}}


In the last years one of the main interests in nuclear physics has been 
focused on the study of halo nuclei, {\it i.e.}, 
weakly bound and spatially extended
systems where one or two particles (generally neutrons) have a high probability
of being at distances larger than the typical nuclear radii 
(see refs.~\cite{HJJ95,Zhu93} for a general review on these nuclei).
The ability to produce secondary beams of halo nuclei opened
new possibilities of investigating their structure. 
Two basic experimental probes involving high energy reactions
have been developed to study halo structure.
The first one is to
measure the momentum distributions of the fragments coming out
after a collision with light stable nuclei~\cite{Kob88,Orr92}.
The second probe treats the analysis of 
Coulomb break-up cross section when the nuclei are incident 
on highly charged targets~\cite{Tan85,BR94,EBB95}. 

The first type of reactions has been treated in detail
in a series of publications by the
group of Aarhus~\cite{GFJ96,GFJ97,GFJ97a,GFJ98}.
Here, the simplest approach to understand halo 
nuclei fragmentation reactions involves the instantaneous removal of one 
of the particles from the few-body halo system. Within this approach, known as
``sudden approximation'', one assumes that the binding system is removed 
without disturbing the motion of the constituent particles. 
This approximation is only
justified for reaction times much shorter than the characteristic time 
for the motion of the particles within the few-body system. 
The sudden approximation has been extensively applied to the study of 
three-body halo nuclei, and in particular to the Borromean systems, 
{\it i.e.}, three-body systems where all two-particle subsystems are 
unbound~\cite{Zhu94,KK94,ZJ95}.
Final interaction between the two non-disturbed spectators
seems to play a crucial role in order to explain the narrow neutron momentum 
distributions measured.
The participant-target interaction was first
described considering only absorption.
Further improvements have been included recently, treating the
interaction between the target and each of the halo particles
by means of a phenomenological optical potential~\cite{GFJ98a}.
The total cross section is then obtained by adding the contributions from all
the participants in the halo nucleus. Processes where two or three halo
particles interact simultaneously with the target are neglected. This is 
consistent with the fact that the model is only accurate for the outer
part of the wave function~\cite{GFJ98,GFJ98a}. 
This means that those geometric configurations
where more than one halo particle get close to the target during the collision
should be excluded. This shadowing effect has been treated in previous works
under different approaches. In the analysis of the Aarhus group the shadowing
is accounted for by excluding the participant wave function inside spheres
around the two spectators~\cite{GFJ98,GFJ98a}.

The second type of probe to study the structure of halo nuclei is by means
of Coulomb elastic break-up reactions with a projectile, composed by a core
and valence neutrons, incident on highly charged targets.
The Surrey group has studied 
in detail elastic scattering of 
halo nuclei from target within the  
``adiabatic'' approach, {\it i.e.}, the intrinsic motion is very slow
compared to the scattering motion~\cite{Joh97,Joh97a,Tos98,Tos98a}. 
Moreover, the
interaction between the projectile and the target is described
considering only the interaction between the core and the target.
This requirement is relevant to Coulomb dominated processes
when the core is charged and the valence particle is neutral. 
In the case
that strong interactions dominate, the above requirement is most likely to be
valid when the number of core nucleons greatly exceeds the number of valence
nucleons~\cite{Joh97}.
Within these approximations, the elastic differential cross section factorizes
into two terms, the cross section for a point-like projectile scattered by the
target, and a form factor that contains the effects of the 
projectile structure.
The range of validity of the adiabatic
approximation is also discussed in~\cite{Joh97a,Tos98} 
concluding that for a pure
strong interaction the adiabatic approach is justified for a given
projectile--target system at sufficiently high energy. On the contrary, in the
case in which the Coulomb interaction dominates
the validity of the ``adiabatic'' approach is questionable at forward
scattering angles.

Apart from the approaches mentioned, several other models have been proposed 
in the
literature, starting from the pioneering work of Bang and Pearson~\cite{Bang},
including eikonal~\cite{Bertsch,Brooke,Barranco},
semi-classical~\cite{Bertulani,Romanelli}, and mixed approaches to
describe direct and sequential break-up~\cite{Zhukov}. 

The paper is organized as follows.
In section 2 we develop the formalism of the uncorrelated scattering 
approximation (USA). Here the basic assumption is 
to neglect the correlation between the fragments in the region where the
interaction with the target is strong. In this situation the orbital angular
momenta and kinetic energies of the fragments are conserved during the
collision process, and this leads to an analytic expression for the S-matrix
of the composite system in terms of the S-matrices of the fragments. 
In section 3 we present a preliminary application of the developed approach
to the case of elastic and break-up deuteron scattering on $^{58}$Ni.
In section 4 the conclusions are presented.

\section{The Uncorrelated Scattering Approximation}

In this section we introduce a new approach to describe the scattering
of a weakly bound nucleus by a heavy target. 
As it will be shown later, the S-matrices to the bound and  break-up
states of the composite system are given 
in terms of the ground state wave function
and the S-matrices corresponding to the interaction of the fragments with 
the target.

The interaction of a composite particle with the target can be expressed as 
the sum of two terms. On one side, an average force acting on the
centre of mass of the projectile, which makes the
projectile to scatter but does not excite or break it. On the other side,
tidal forces that make the projectile rotate, excite or break  up. Then,
when a composite particle scatters from a target there are two opposite
effects: i) the interaction between the fragments tending to
keep the fragments bound, and ii) the tidal forces tending to break
the system.
In order to simplify our discussion, we assume that the mass of the
target is much larger than the masses of the fragments. The Hamiltonian
can be written then as

\begin{eqnarray}
\label{eq:Hmod}
H&=&\frac{\vec{P}^{2}}{2M}+\frac{\vec{p}^{2}}{2m}+v_{AB}(r)
+v_{AT}(R_{AT})+v_{BT}(R_{BT})\\
 &=&\frac{\vec{P}_{A}^{2}}{2m_A}+\frac{\vec{P}_{B}^{2}}{2m_B}+v_{AB}(r)
+v_{AT}(R_{AT})+v_{BT}(R_{BT})
\end{eqnarray}
where $M=m_A+m_B$ and $m=m_A m_B/(m_A+m_B)$. In this model, the interaction
between projectile and target can be written as the sum of a 
folding potential, $v_F(R)=\langle\phi_0|v_{AT}+v_{BT}|\phi_0\rangle$, 
which does not 
affect the internal structure of the projectile, and a tidal potential,
$v_T(R, r)= v_{AT}(R_{AT})+v_{BT}(R_{BT})-v_F(R)$, which tends to break the 
projectile. The function $|\phi_0\rangle$ describes the intrinsic ground state
of the projectile.
Note that for large distances $R$, the tidal forces, coming from the gradient
of $v_T$,
are negligible compared to the force between the fragments, coming from 
the gradient of $v_{AB}$,
that tend to keep them bound. Hence,
it is reasonable to ignore the tidal forces for large distances.

On the contrary, for small distances $R$, tidal forces can be large.
In this case a reasonable approach is
to ignore the force between the fragments,
substituting the potential $v_{AB}$ 
by a suitable constant $\bar v$.
For very tightly bound systems, tidal forces may not be strong enough to 
overcome the force between the fragments for any distance $R$. 
For these systems, 
scattering will be predominantly elastic and governed by the folding 
potential. However, for weakly bound systems there will be a critical distance
$R_m$ below which tidal forces overcome the force between the fragments.
The distance $R_m$ can be associated to an angular momentum $L_m$, so that
$R_m$ is the turning point of the wave-function, fulfilling
\begin{equation}
{L_m (L_m+1) \over 2 M {R_m}^2} + v_f(R_m) = E - \epsilon_0 \,\,\, .
\end{equation}
Note that for $L>L_m$, tidal forces are not very important because the
turning point is beyond $R_m$. On the contrary, for $L<L_m$ tidal forces will
be important, and the correlation between the fragments may be neglected.

Let us consider the situation in which tidal forces can be neglected. 
Thus, the Hamiltonian $H$, approximated by $H^F$, 
can be decomposed as follows,
\begin{eqnarray}
H^F &=& h_r + h_R \\
h_r &=&\frac{\vec{p}^{2}}{2m} + v_{AB}(r) \\
h_R &=&\frac{\vec{P}^{2}}{2M} + v_F(R) \,\,\, .
\end{eqnarray}
The eigenstates of $H^F$ for a total energy $E$ can be 
expanded in terms of
products of eigenstates of the internal Hamiltonian $h_r$ corresponding to
energies $\epsilon_n$, times 
eigenstates of $h_R$ corresponding to energies $E-\epsilon_n$. 
We make use of a discrete and finite
basis of $N$ normalizable states of the 
relative motion of the fragments. These basis states include the
bound states of the projectile and the resonant states of the continuum.
Diagonalizing the internal Hamiltonian $h_r$
in this basis, one obtains the eigenstates $|nIM\rangle$ 
with internal energies $\epsilon_n$. Thus, the
energy of the relative motion of the projectile and target in the asymptotic
region is $E_n=E-\epsilon_n$.
The states that correspond to energies $E_n<0$ do not
contribute to the wave function asymptotically and, therefore, we 
restrict our basis space to $E_n>0$.
The states $|nIM\rangle$ are characterized by a given angular
momentum $I,M$. Thus, we can write
\begin{equation}
\langle\vec r|nIM\rangle = \phi_n(r) Y_{IM}(\hat r)
\end{equation}
\begin{equation}
\langle\vec p|nIM\rangle = \tilde \phi_n(p) Y_{IM}(\hat p)
\end{equation}
in coordinate and momentum space, respectively. 

For the purpose
of defining scattering magnitudes, let us consider the Hamiltonians free from
the interaction with the target $H^{0}$, where 
$H^F=H^{0}+v_F(R)$. The regular solutions of $H^{0}$ are 
characterized by a total energy $E$, orbital 
angular momentum $L$, internal angular momentum of the fragments $I$, 
total angular momentum $J$ and internal energy $\epsilon_n$: 
\begin{equation}
|\Psi_{nILJM_J}^{0}(E)\rangle= {\cal J}_{L}(P_{n}R)|n(LI)JM_J\rangle ,
\end{equation}
where ${\cal J}_L$ represents a regular wave function of the 
free Hamiltonian that is just proportional to 
a Bessel function $j_L(P_nR)$, whereas
$|n(LI)JM_J\rangle$ is the channel wave function in 
which the internal state with angular momentum
$I$ is coupled to the relative angular momentum $L$ to produce the
total angular momentum $J,M_J$. The momentum
associated to the relative motion is given by $\vec{P}_n^2 /2M
= E - \epsilon_n$. If one constructs wave packets out of this wave function,
one will have incoming waves for $t \to -\infty$,  and outgoing waves for 
$t \to +\infty$. The  scattering
in $H^{F}$ is such that
the incoming waves will be unmodified, while the outgoing waves
will be affected by the S-matrix due to the folding potential
$S_F(L,E_n)$, which will be diagonal in the channel basis.

Let us now neglect the correlation between the projectile fragments.
Then, the interaction $v_{AB}$ is replaced by a constant $\bar v$.
The total Hamiltonian can be written in terms of two non-interacting 
Hamiltonians
\begin{eqnarray}
\bar H &=& h_A + h_B + \bar v \\
h_A &=&\frac{\vec{P}_A^{2}}{2m_{A}} + v_{AT}(R_{AT}) \\
h_B &=&\frac{\vec{P}_B^{2}}{2m_{B}} + v_{BT}(R_{BT})\,\,\, .
\label{eq:Hbar}
\end{eqnarray}
The eigenstates of $\bar H$ corresponding to an energy $E$
can be expanded in terms of the product of eigenstates of $h_A$ and $h_B$, 
such that $E=E_A+E_B+\bar v$. 
Given the adequate boundary conditions, it is straightforward to
solve the scattering problem for the Hamiltonian $\bar H$. We consider
the Hamiltonian free from interactions with the target $\bar H^0$, so that
$\bar H = \bar H^0 + v_{AT}(R_{AT}) + v_{BT}(R_{BT})$. A solution of
this Hamiltonian is given by the product of regular wave functions in the 
co-ordinates
$R_{AT}$ and $R_{BT}$, 
characterized by angular momenta 
$L_A, M_A$ and $L_B, M_B$, and energies $E_A$ and $E_B$.
\begin{equation}
|\Psi^0_{L_AM_AL_BM_B}(E)\rangle 
= {\cal J}_{L_A}(P_AR_{AT}){\cal J}_{L_B}(P_BR_{BT})  
|L_AM_AL_BM_B\rangle .
\end{equation}
Then,  constructing a wave-packet, we find that for 
$t\to -\infty$, the wave function is given by the product of incoming
wave functions, while for $t\to +\infty$ it is given by 
the product of outgoing wave functions. Cross terms containing the product of 
an incoming wave on one co-ordinate and an outgoing wave on the other are 
cancelled for $t \to \pm\infty$.
Then, if we switch on the interactions
$v_{AT}+v_{BT}$,  
the incoming part is unaffected, while the outgoing
part gets  multiplied by the product of the elastic S-matrices generated
by each potential. This means that the three-body S-matrix for the 
Hamiltonian $\bar H$ is diagonal in the basis characterized by the linear 
momenta and angular momenta of each fragment, and is given by the product 
of the S-matrices of each fragment $S(L_A,E_A) S(L_B,E_B)$. 

A basic point in order to deal with the matching is to realize that
the wave functions 
of $\bar H^0$ 
can be characterized by the hyper-angular
momentum $K$. In a basis of hyper-spherical harmonics, 
the wave functions obtained in the absence of interactions can be written
as
\begin{equation}
|\Psi^0_{KILJM_J}(E)\rangle 
= {\cal J}_{K} ({\cal P}{\cal R}) |K(LI)JM_J\rangle ,
\end{equation}
where ${\cal J}_{K}(x)$, that is proportional to the Bessel function
$J_{K+2}(x)$, is a regular solution of free three-body problem in terms
of the hyper-radius 
${\cal R}$, given by ${\cal R}^2=R^2+r^2 m/M$, and the hyper-momentum 
${\cal P}$, given by  ${\cal P}^2/2M = E - \bar v$. 
The wave function $|K(LI)JM_J\rangle$ can be written in terms of the 
hyper-angle 
$\alpha$ that defines
the ratio of $p$ to ${\cal P}$, {\it i.e.}, 
$\sin \alpha = (\sqrt{M/m}) p/{\cal P}$. 
Explicitly, 
\begin{equation}
|K(LI)JM_J\rangle 
= \int_0^{\pi/2} d \alpha f^{LI}_K(\alpha) |\alpha (LI)JM_J\rangle ,
\end{equation}
with $f^{LI}_K(\alpha)$ a function given in terms of the Jacobi 
polynomials (see appendix). 
The hyper-angular momentum $K$ provides an upper bound for $L$ and
$J$, {\it i.e.}, $J\le L+I \le K$. 
%
%
 Thus, if we take a value
of $K$ given by $K_m=L_m$, we can argue that for $K>K_m$ tidal forces 
are less important than the forces between the 
fragments. On the contrary, when $K\le K_m$ the forces 
between the fragments will be small compared to the tidal forces.
The relative importance of tidal forces compared to the forces between the 
fragments depends obviously on the values of $R_{AT}, R_{BT}$ and $r$. 
However, within the USA approach, such relative importance
between both types of forces is basically determined by the
value of $K$.

The  Uncorrelated Scattering Approximation (USA) uses the expansion of the 
scattering
wave function in terms of the hyper-angular momentum $K$. 
Then, the Hamiltonian $H$ is approximated by $H^{F}$ 
for the components of the wave function such that
$K>K_m$. Thus, tidal forces
are ignored and the scattering is governed by the folding potential.
This means that no excitation or break-up of the 
projectile occurs in these components.
On the contrary, for $K\le K_m$ the Hamiltonian $H$
is approximated by $\bar H$. Then, the correlation between the particles is 
ignored. It is very important to realize that the USA formulates different
approximations for $H$ in terms of the value of $K$, and not in terms of $R$.
Thus, $\bar H$ ($H^{F}$) is the approximate expression of $H$ for any $R$,
provided that $K\le K_m$ ($K>K_m$).

It is important to formulate the USA to ensure that the
interaction does not couple states with $K\le K_m$ to states with $K>K_m$.
In order to do that, let $P$ be the projector on the states with
$K\le K_m$ and $Q$ the projector on the rest of states. The full Hamiltonians
$H$ can be expressed as $HP + HQ$. The USA implies that the term $HP$ is
approximated by $P \bar H P$, while $HQ$ is approximated by $Q H^{F}$.
This ensures that the time evolution of a state $|i\rangle$ is given
by the sum of the evolution of $P|i\rangle$ and that of $Q|i\rangle$, which
remain mutually orthogonal.

We can now study what our approximation implies regarding the S-matrix.
We start with a regular solution of $H$ and then we construct a wave packet.
For $t \to -\infty$ the wave packet will be
characterized by an incoming wave function 
times an internal state given by the ket $|i\rangle=|nLIJM_J\rangle$.
The wave-packet  for
 $t \to +\infty$  will be a product of outgoing waves times
a combination of states $|f\rangle=|n'L'I'J'M_J'\rangle$, multiplied by 
certain coefficients, which are the matrix elements of the 
S-matrix operator between the states $\langle f|$ and $|i\rangle$.
 Then, 
\begin{equation}
\langle  f|S|i\rangle = \langle f|SP|i\rangle + \langle f|SQ|i\rangle.
\end{equation}
Within the USA model, the Hamiltonian $H$ is replaced by $H^{F}$ 
when referred to states with $K>K_m$. This implies that the 
operator $SQ$ can be approximated by 
$S_F Q$, where $S_F$ is
a c-number given by the elastic S-matrix for the calculation 
involving the folding potential. For $K\le K_m$
the Hamiltonian $H$ can be substituted for $\bar H$, implying that
the operator $SP$ can be approximated by $\bar S P$, where $\bar S$ is the
S-matrix for the Hamiltonian $\bar H$. As we will see in next section,
$\bar S$ can be expressed in terms of the product of S-matrices of the two
particles $A$ and $B$, and it remains in the space of states with $K\le K_m$.
Thus, $SP \simeq P \bar S P$. Finally, we can write
\begin{equation}
\label{eq:SplusDelta}
\langle f|S|i\rangle \simeq S_F \delta_{f,i} + 
\langle f|\Delta S|i\rangle , 
\end{equation}
where $\Delta S = P (\bar S - S_F) P$. Therefore, 
within the Uncorrelated Scattering Approximation, the
S-matrix is given by the sum of two terms: 
the S-matrix coming from the folding model,
which contributes only to the elastic scattering, and  a correction term
that affects only to the components with $K \le K_m$, and which contains 
all the excitation and break-up effects. This term
is given by the difference between the S-matrices of the two 
uncorrelated fragments
and the S-matrix from the folding model.  

\subsection{Boundary conditions}


Let us proceed now to describe the boundary conditions. Consider
a regular solution of $H^{0}$,  
characterized by a total energy $E$, orbital 
angular momentum $L$, internal angular momentum of the fragments $I$, 
total angular momentum $J$ and internal energy $\epsilon_n$: 
\begin{equation}
\label{wasym}
|\Psi_{nILJM_J}^{0}(E)\rangle= {\cal J}_{L}(P_{n}R)|n(LI)JM_J\rangle .
\end{equation}
The state $|n(LI)JM_{J}\rangle$ can be written explicitly as
\begin{equation}
|n(LI)JM_{J}\rangle=\int_{0}^{\infty}p^2 dp \tilde{\phi}_n(p)|p (LI) JM_{J}\rangle .
\end{equation}
The channel wave function $|n(LI)JM_J\rangle$ 
can be projected with the operator 
$P$, extracting the components with $K\le K_m$. 
Thus, we have
\begin{equation}
P|n(LI)JM_J\rangle = \sum_{K=L+I}^{K_m} \langle K |n\rangle _{LI} 
|K(LI)JM_J\rangle ,
\end{equation}
where the overlap is given by
\begin{equation}
\langle K|n\rangle_{LI} =
\int_0^{p_m} dp p \left[{d\alpha(p)\over dp}\right]^{1/2} 
f^{LI}_K(\alpha(p))^* 
\tilde \phi_n(p).
\end{equation}
 
Thus, the asymptotic regular wave function, projected by $P$ and written 
as an eigenstate of $\bar H^0$, becomes   
\begin{equation}
\label{wasym1}
P|\Psi_{nLIJM_J}^{0}(E)\rangle = \sum_{K} \langle K|n\rangle_{LI}
|\Psi_{KLIJM_J}^{0}(E)\rangle \,\,\, 
\end{equation}
where
\begin{equation}
|\Psi_{KLIJM_J}^{0}(E)\rangle = 
{\cal J}_{K}({\cal P}{\cal R}) |K(LI)JM_J\rangle \,\,\, .
\end{equation} 

Using the Raynal-Revai transformation~\cite{Raynal},
the incident wave function can be expressed in terms of the
angular momenta associated to the coordinates $R_{AT}, R_{BT}$,
\begin{equation}
|K(LI)JM_J\rangle = \sum_{L_A L_B} \langle L_A L_B|L I\rangle_{JK} 
|K(L_AL_B)JM_J\rangle ,
\end{equation}
with $J\le L_A+L_B \le K$. Thus, we can write
\begin{equation}
|\Psi_{KLIJM_J}^{0}(E)\rangle = \sum_{L_A L_B}
\langle L_A L_B|L I\rangle_{JK}
|\Psi_{KL_AL_BJM_J}^{0}(E)\rangle \,\,\, .
\end{equation}
Note that the Raynal-Revai coefficient vanishes for $K < L+I$ 
or $K< L_A+L_B$. 
The state $|\Psi_{KL_AL_BJM_J}^{0}(E)\rangle$ 
is a regular solution of $\bar H^{0}$ for specific 
values of $L_A$, $L_B$ and $K$
\begin{equation}
|\Psi_{KL_AL_BJM_J}^{0}(E)\rangle = 
{\cal J}_{K}({\cal P}{\cal R}) |K(L_AL_B)JM_J\rangle \,\,\, .
\end{equation} 
The angular momenta $L_A$ and $L_B$ are separately
conserved in the scattering process due to $\bar H$. 
However, the hyper-angular momentum
$K$, which is a good quantum number for $\bar H^{0}$,
 is no longer conserved by $\bar H$. Thus, one can
proceed by using a new basis that keeps $L_A$ and $L_B$
as quantum numbers, but replaces $K$ with other quantum number which
is conserved by $\bar H$. Note that this Hamiltonian~(\ref{eq:Hbar})
conserves the energy, and hence the
asymptotic momentum of each particle separately. In the appendix we show
how to transform the states characterized by the values
of $K$ up to $K_m$ into states that have, approximately, 
a defined value of the momentum of each particle. This transformation
is achieved in terms of the Momentum Localized States (MLS)
$|\ell (L_A L_B) J M_J\rangle$. These states depend
on the momenta $P_A$ and $P_B$ which are strongly
peaked around the values $P_A^\ell={\cal P}\sqrt{M_A/M}\cos(\beta_{\ell})$ and
$P_B^\ell={\cal P}\sqrt{M_B/M}\sin(\beta_{\ell})$, respectively. 
The energies are given by $E_A^\ell=(E-\bar v)\cos^2(\beta_{\ell})$
and $E_B^\ell=(E-\bar v)\sin^2(\beta_{\ell})$.
The relation between the localized states and the original states is given 
by means of an orthogonal transformation,
\begin{equation}
|K (L_A L_B) J M_J\rangle = \sum_{\ell=1}^{n_\ell} 
\langle \ell|K\rangle_{L_A L_B}
|\ell (L_AL_B)JM_J\rangle ,
\end{equation}
where the number of momentum localized states, $n_\ell$, coincides 
with the number
of states with definite $K$, $n_\ell=[(K_m-L_A-L_B)/2]+1$.
The coefficients of the transformation are analytic expressions given
in the appendix.
Then, we can write
\begin{equation}
\label{wasym2}
|\Psi_{KL_AL_BJM_J}^{0}(E)\rangle = \sum_{\ell=1}^{n_\ell}
\langle \ell|K\rangle_{L_A L_B}
|\Psi_{\ell L_AL_BJM_J}^{0}(E)\rangle \,\,\, .
\end{equation}
The  state $|\Psi_{\ell L_AL_BJM_J}^{0}(E)\rangle$
corresponds to a regular wave function in which the two particles $A$ and $B$ 
have linear momenta with narrow distributions around
$P_A^\ell$ and $P_B^\ell$, and angular momenta $L_A$ and $L_B$,
respectively. 
If we define 
\begin{equation}
\langle\ell L_A L_B|nLI\rangle_J=\sum_{K=L_A+L_B}^{K_m} 
\langle K|n\rangle_{LI}
\langle L_A L_B|L I\rangle_{JK} \langle\ell|K\rangle_{L_A L_B} ,
\end{equation}
which is a coefficient that depends on the bound wave functions and on analytic
transformation coefficients, we can write finally,
\begin{equation}
\label{wasymf}
P|\Psi_{nILJM_J}^{(0)}(E)\rangle = \sum_{L_A L_B}
\sum_{\ell=1}^{n_\ell} \langle\ell L_A L_B|nLI\rangle_J 
|\Psi_{\ell L_AL_BJM_J}^{0}(E)\rangle  .
\end{equation}
Note that this transformation relates the asymptotic states 
that define the
boundary condition of the scattering problem, with the states for which each 
particle has a well defined angular and linear momentum.
From this expression, we can construct the incoming and outgoing waves just
by making the adequate wave-packets. 
For $t \to -\infty$, eq.~(\ref{wasymf}) relates the 
incoming parts of 
$P|\Psi_{nILJM_J}^{(0)}(E)\rangle$ and 
$|\Psi_{\ell L_AL_BJM_J}^{0}(E)\rangle $,  
while for $t \to +\infty$, it relates the outgoing parts.

\subsection{S-matrix to bound and resonant break-up states}

Within the Uncorrelated Scattering Approximation, 
the two particles scatter independently inside the 
interaction region where the full Hamiltonian, projected on values
$K \le K_m$, $P H P$, is replaced by $P \bar H P$. 
The S-matrix is simply expressed in
a basis of momentum localized states $|\Psi_{\ell L_AL_BJM_J}^{0}(E)\rangle $. 
A wave packet of these states at $t \to -\infty$ evolves according to 
$P \bar H P$ to give
for $t \to +\infty$, the product of the S-matrices 
$S_A(L_A,E_A^\ell) S_B(L_B,E_B^\ell)$ times the wave-packet. Note that in
writing this expression, one substitutes the
narrow energy distributions of $E_A$ and $E_B$ of the MLS state for their 
central values $E_A^\ell, E_B^\ell$.

The matrix elements of $P\bar S P$ in the channel basis $|nLIJ\rangle$ can be
evaluated considering the transformation (\ref{wasymf}),
{\setlength\arraycolsep{2pt}
\begin{eqnarray}
\langle n'I'L'J|P \bar S P|nILJ\rangle & = &  
\sum_{L_A L_B}\sum_{\ell=1}^{n_\ell} 
\langle \ell L_A L_B|nLI\rangle_J \langle n'L'I'|\ell L_A L_B\rangle_J \nonumber \\
& \times & S_A(L_A,E_A^\ell) S_B(L_B,E_B^\ell),
\end{eqnarray}}
with $J\le L_A+L_B \le K_m$.

In order to evaluate the matrix elements for $P S_F P$ one should note that 
the operator $S_F$ is a function of the orbital angular momentum
$L$ and the energy of relative motion  $E_n$. It
conserves the orbital angular momentum $L$ and the internal angular momentum
$I$, and is independent on $K$. Moreover, the operator $P$ conserves
$L$ and $I$ and projects on $K\le K_m$. Thus, we can write
\begin{equation}
\langle n'I'L'J|P S_F P|nILJ\rangle = \delta_{I'I}\delta_{L'L} S_F(L,E_n)
\sum_K^{K_m} \langle n'|K\rangle_{IL} \langle K|n\rangle_{IL},  
\end{equation}
that can be also expressed in the form,
\begin{equation}
\langle n'I'L'J|P S_F P|nILJ\rangle =
\sum_{L_A L_B} \sum_{\ell=1}^{n_\ell}
\langle \ell L_A L_B|nLI\rangle_J \langle n'L'I'|\ell L_A L_B\rangle_J 
S_F(L,E_n),
\end{equation}
where we have used orthogonality properties.

Then, the final expression for the S-matrix in the Uncorrelated Scattering
Model results
\begin{equation}
\label{eq:S_USA}
\langle n'I'L'J| S |nILJ\rangle = \delta_{n'n} \delta_{I'I} 
\delta_{L'L} S_F(L,E)
+ \langle n'I'L'J| \Delta S |nILJ\rangle,
\end{equation}
with
{\setlength\arraycolsep{0pt}
\begin{eqnarray}
\langle n'I'L'J|\Delta S|nILJ\rangle & = &  \sum_{L_A L_B} 
\sum_{\ell=1}^{n_\ell}
\langle \ell L_A L_B|nLI\rangle_J \langle n'L'I'|\ell L_A L_B\rangle_J 
\,\,\,\,\, \nonumber \\
& \times &
\left\{
S_A(L_A,E_A^\ell) S_B(L_B,E_B^\ell) - S_F(E,L)\right\}.\,\,\,\,\,\,\,
\end{eqnarray}}
This expression is valid for elastic scattering, inelastic scattering
to bound states and break-up to resonant states in the continuum. We also
notice that the expression (\ref{eq:S_USA}) can be interpreted 
as the matrix element of the operator $S_{F}+\Delta S$, 
with $\Delta S=P(\bar{S} -S_{F})P$, between the initial 
and final internal states.


\subsection{Partial non-resonant breakup cross section}

In the previous section we have derived expressions for the S-matrix
elements corresponding to final states, bound or resonant, that can be 
represented by normalizable wave functions. They are given in terms of
the matrix elements of the operator \( \Delta S \) that is naturally 
described in the MLS basis \( |\ell (L_{A}L_{B})J\rangle  \), in which this
operator is diagonal. Thus, only the overlap between
the wave function of the final state and the states in the MLS basis
is required to be known.
The same procedure could be applied to calculate break-up to non-resonant
continuum states. Provided that the corresponding wave functions are known in 
momentum representation, the overlap can be obtained.

In this section we do not calculate the expressions for the
break-up to specific states in the continuum as they should depend on 
the detailed continuum states wave functions considered. 
Instead, we derive 
closed expressions for the non-resonant break-up cross sections, integrated
over all the possible values of the energies of the fragments, but 
characterized by a certain angular momentum of the fragments \( I' \) and of
the relative motion \( L' \).

Making use of the completeness
relation for the internal eigenstates, it is possible to 
derive a closed expression
for the partial breakup cross section leading from the initial bound state
$|nLIJ\rangle$
to all the final non-resonant continuum states
characterized by the set of angular momenta \( \{L',I',J\} \). We denote
this cross section by \( \sigma_J^{bu}(nLI\to L'I') \).
The details of the derivation are given in \cite{mitesis} and will be published
elsewhere. 
In this situation the final expression for the
partial breakup cross section within the USA is given by

\begin{eqnarray}
\sigma_J ^{bu}(nLI\to L'I') & = & \frac{\pi }{P^{2}_{0}}(2L+1)\Big {\{}\sum _{K}|\langle K(L'I')J|\Delta S|nLIJ\rangle |^{2}\nonumber \\
 &  & \, \, \, \, \, \, \, \, \, \, \, \, \, \, 
-\sum_{n'}|\langle n'L'I'J|\Delta S|nLIJ\rangle |^{2}\Big {\}},\label{eq:xsebuLIJ_USA} 
\end{eqnarray}
where \( P_{0} \) is the asymptotic incident momentum of the projectile. This
expression has a simple interpretation. The first term is the cross section
induced by the operator \( \Delta S \) to all the states labeled by the angular
momenta \( K,L',I',J \), that include the contribution of the bound and 
resonant states,
which are explicitly subtracted by the second term. The summation with 
respect to $K$ is extended to all the values between $L'+I'$ and $K_m$.

It is also possible to obtain a compact expression for the breakup cross section
corresponding to a total angular momentum \( J \), \( \sigma ^{bu}_{J}(nLI) \).
This is achieved upon summation of \(\sigma_J ^{bu}(nLI\to L'I')\) 
on the angular
momenta \( L' \) and \( I' \) and taking into account the completeness property
of the states \( |K(L'I')JM_{J}\rangle  \). This leads to the close expression

\begin{equation}
\label{eq:xsecbuJ_USA}
\sigma ^{bu}_{J}(nLI)=\frac{\pi }{P^{2}_{0}}(2L+1)
\Big {\{}\langle nLIJ|(\Delta S)^{+} \Delta S|nLIJ\rangle -
\sum_{n'L'I'}|\langle n'L'I'J|\Delta S|nLIJ\rangle |^{2}\Big {\}}.
\end{equation}

Then, within the USA, the non-resonant breakup cross section for a given total 
angular momentum is calculated as the dispersion of the operator 
\( \Delta S \) in the ground state of the projectile, subtracting the 
contribution of the other bound and resonant states.

\section{Application to the \protect\( d\protect \) + \protect\( ^{58}\protect \)Ni
reaction}

In this section we apply
the uncorrelated scattering approximation to the analysis of
elastic and breakup scattering of \( d \) by \( ^{58} \)Ni. 
Though the USA is expected to
work better for more loosely bound projectiles such as \( ^{8} \)B
or \( ^{11} \)Be, for which the correlations between the fragments are
weaker than for the deuteron, we start
studying the case of the deuteron because this is a
much better known system for which numerous calculations
and experimental data already exist. 

The reaction \( d \) + \( ^{58} \)Ni has been extensively studied by the Kyushu
group by means of Continuum Discretized Coupled Channel Calculations (CDCC)
\cite{Yah86,Ise86,Aus87}. It has also been used as a test case of the adiabatic
approximation \cite{Ron70} and the Glauber multiple-scattering theory \cite{Yab92}.
All these approaches predict an important effect of the coupling to the
breakup channels that results in a significant departure of 
their predictions compared to the
folding model calculation. This is a characteristic phenomenon of reactions
involving halo nuclei. Therefore, some of the conclusions arising from the
analysis of reactions with deuterons can be also extended 
to the case of exotic nuclei. 

We first analyze the elastic scattering data at 80 MeV. As already mentioned,
the calculation of the elastic \( S \)-matrix elements within the
USA requires the following ingredients:

i) The internal wave function of the deuteron. Within 
the USA this wave function
enters in both the folding potential and the coefficients
\( \langle n|K\rangle _{LI} \) appearing in eq.~(\ref{eq:S_USA}). We adopt
in this work a simple model of the deuteron which results from the assumption
that the proton-neutron potential is separable in momentum space \cite{Cris98}.
In this model the S-wave component of the deuteron ground state 
is described in momentum
space by the simple analytic expression

\begin{equation}
\label{eq:wfcris}
\tilde{\phi }_{0}(p)=N\frac{\exp \left( -p^{2}/2mC\right) }{p^{2}+2mB},
\end{equation}
where \( \hbar /\sqrt{mC} \) is related to the range of the proton-neutron
interaction, \( B \) is the binding energy of the deuteron (\( B=2.22 \) MeV)
and \( N \) is a normalization constant. We neglect the small D-wave component
of the ground state wave function and the proton and neutron intrinsic spins.

ii) The second ingredient of the USA refers to 
the two-body \( S \) matrices for the
constituents. In the case of the deuteron, the proton-target 
and neutron-target $S$-matrices ($S_p,S_n$) are required for 
values of the angular momenta in the interval 
\( 0\leq L_{p},L_{n}\leq K_{m} \),
and values of the energies determined 
by the momentum localized states, which lie in the
range \( 0\leq E_{p},E_{n}\leq E-\bar{v} \). These \( S \) matrices have been
calculated by means of optical potentials as it is done in the CDCC, adiabatic
and Glauber calculations. In these formalisms the proton and neutron optical
potentials are evaluated at half of the incident deuteron energy and so, the
energy dependence of the optical potential parameters is neglected. 
This approximation
is based on the assumption that the proton and neutron move approximately with
the same velocity of the deuteron center of mass, and the dispersion around
this value is small. Those configurations for which one of
the fragments carry the whole available energy 
must be highly suppressed. Within the USA this fact is explicitly included
in the coefficients, \( \langle n(LI)J|\ell L_{A}L_{B}\rangle  \), that can
be physically regarded as the amplitude probability of having the constituents
of the projectile with angular momenta \( L_{A} \) and \( L_{B} \), 
and energies
\( E^{\ell }_{A} \) and \( E^{\ell }_{B} \) within a state characterized by
the relative angular momentum \( L \), internal momentum \( I \) and total
angular momentum \( J \). In fact, these coefficients favour 
those configurations
for which each one of the particles carries half of the incident 
angular momentum
and half of the available energy, according to a classical picture. It is important
to note that, within the USA, the energy dependence of the optical potential
is naturally taken into account by solving the proton-target and neutron-target
Schr\"odinger equations at definite scattering energies. In particular, we
perform the calculations with the optical potentials of Ref.\ \cite{BG69}. 

The USA also requires the introduction of
two parameters, the average potential, \( \bar{v} \), and the cut-off 
hyper angular momentum, \( K_{m} \). These parameters have
a clear physical interpretation and so they could be set to some reasonable 
values, or otherwise fitted to improve the agreement with exact calculations. 
Nevertheless, we keep in mind that our purpose in developing the model is to 
apply it to weakly bound systems for which \( K_{m} \) must be large and 
\( \bar{v} \) small. Then, we have adopted the simplest approach. 
We take \( K_{m} \) large enough to achieve convergence for the S-matrix 
elements. This means that the correlations between the fragments can be 
neglected even at large distances. For the average potential we take 
 \( \bar{v} \)=0. 
Several test calculations have revealed a weak dependence on this parameter.
Hence its choice does not affect significantly the results.

\begin{figure}
{\par\centering \resizebox*{0.6\textwidth}{!}{\includegraphics{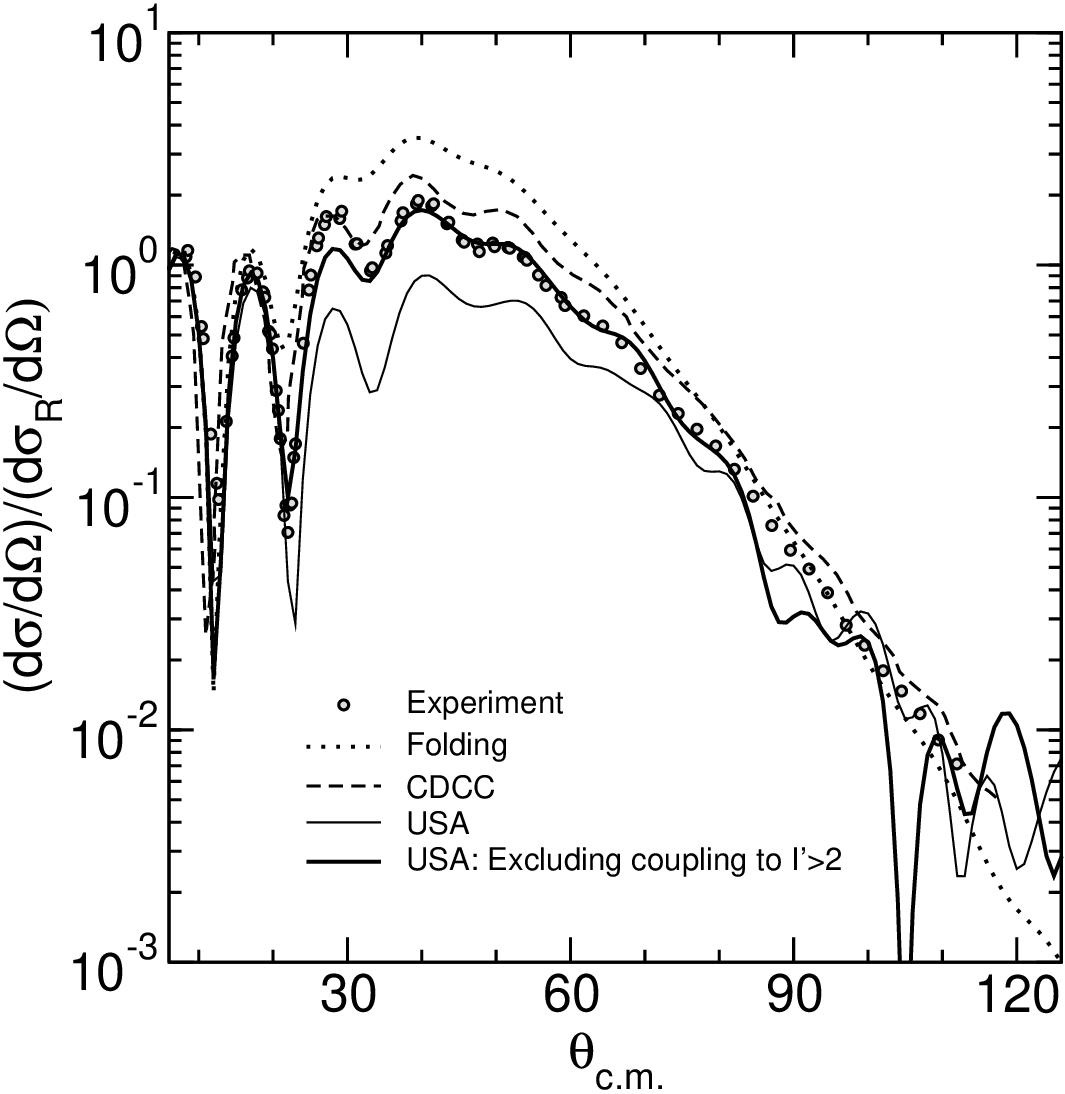}} \par}

\caption{\label{fig:dni_e80_elxsec}Elastic differential cross sections angular distributions
(as ratio to Rutherford) for \protect\( d\protect \) + \protect\( ^{58}\protect \)Ni
scattering at 80 MeV. The dotted, dashed and thin solid line correspond, respectively,
to the folding, CDCC and full USA calculations. 
The thick solid line is the restricted USA calculation which contains only 
the effect of $I'=0,2$
breakup states. Experimental data \cite{Ste83} are given
by circles.}
\end{figure}

The calculated angular distribution of the elastic differential 
cross section, 
divided by the Rutherford cross section, is plotted in 
Fig.~\ref{fig:dni_e80_elxsec}. Experimental data are represented by circles.
The dotted line refers to the folding calculation, which clearly overestimates
the cross section at intermediate angles. The dashed line is the result of
the CDCC calculation taken from \cite{Yah86} that, compared with the folding
model, predicts a significant reduction of the cross section at intermediate
angles. This is a consequence of the coupling to breakup channels. 
The solid line corresponds to the USA calculation with the cutoff 
hyper angular momentum \( K_{m} \)=30, for which convergence of the 
S-matrix elements is achieved. As shown, experimental
data are not accurately reproduced by the full USA calculation that 
overestimates the  reduction of the cross section with respect to 
the folding model due to break-up effects.

\begin{figure}
{\par\centering \resizebox*{0.8\textwidth}{!}{\rotatebox{-90}{\includegraphics{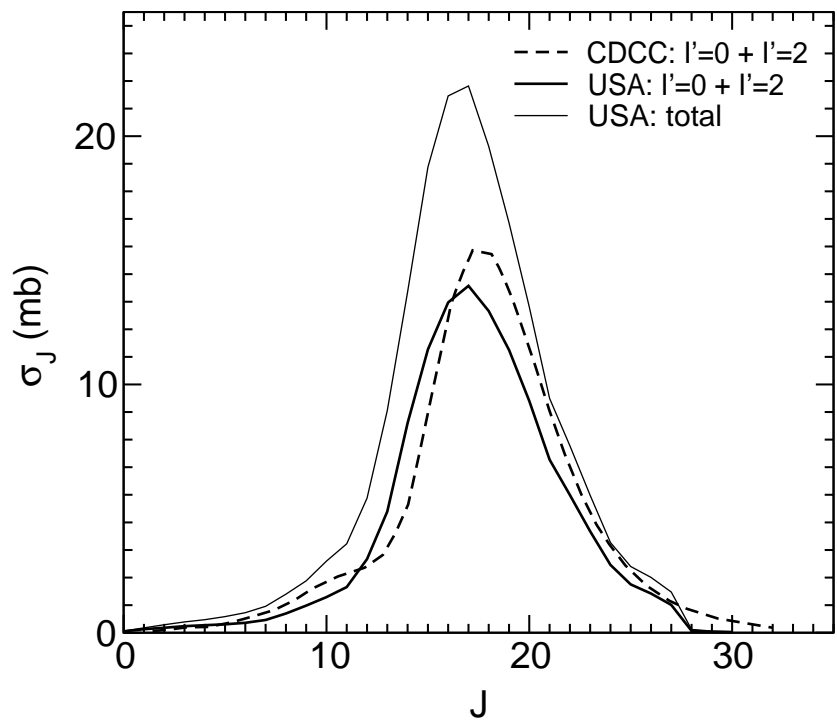}}} \par}

\caption{\label{fig:dni_e80_xsecbuJ}Partial breakup cross section, as a function of
the angular momentum. The thin solid line is total breakup cross section predicted
by the USA calculation (eq.~\ref{eq:xsecbuJ_USA}). The thick solid line corresponds
also to the USA calculation, but including only the S and D continuum channels.
The dashed line is the analogous calculation in the CDCC approach.}
\end{figure}

In order to provide an explanation of this result we investigate the
partial breakup cross section, \( \sigma ^{bu}_{J} \). Within the USA, this
quantity can be easily evaluated without any explicit description 
of the continuum states, by means of eq.~(\ref{eq:xsecbuJ_USA}) that
only requires the introduction
of the ground state wave function. In Fig.~\ref{fig:dni_e80_xsecbuJ} 
we show the distribution
of \( \sigma ^{bu}_{J} \) versus the total angular momentum \( J \) calculated
in the USA (thin solid line). This result is compared with the CDCC 
calculation (dashed line) performed within the subspace 
\( I' \)=0, 2. Thus,
in order to enable a meaningful comparison between both approaches, we have
also calculated within USA,
the contribution
to the breakup cross section due to the S (\( I' \)=0) and 
D (\( I' \)=2) components.
The result, plotted in Fig.\ \ref{fig:dni_e80_xsecbuJ} by the thick solid line,
shows a fairly good agreement with the CDCC 
calculation (dashed line) \cite{Ise86,Yah86}.
Apart from the agreement in the overall magnitude, the angular momentum
dependence is also accurately reproduced, 
including the surface-peak nature of the elastic breakup process.  

From the results for the partial breakup cross section (Fig.~2), it is
clear that the USA predicts an important contribution to the breakup coming
from continuum states with internal angular momenta \( I'>2 \). 
Austern {\it et al.}~\cite{Aus87} have studied the convergence 
of the CDCC calculation for this
reaction with respect to the cut-off internal angular momentum. 
Their calculations
reveal a small contribution to the breakup cross section coming from breakup
channels with \( I'>2 \). For example, within the model 
space \( I' \)=0,1,2,4,6,
for which a good convergence of the solution is achieved, the breakup cross
section associated to \( J \)=17 is 17.27 mb, whereas the USA predicts a value
of 22 mb. We interpret this discrepancy as a consequence of 
the basic approximation
involved in the USA, {\it i.e.,} to neglect the correlations between 
the constituents
in the scattering process. In this sense, our treatment is opposite 
to the adiabatic approximation, in which the internal coordinate is 
assumed to be frozen during
the collision, keeping the constituents strongly correlated,
and thus avoiding the breakup to high angular momentum states. 

This effect could also explain the discrepancy encountered for the differential
elastic cross section. In order to provide a numerical assessment of this 
hypothesis we have performed a new calculation for the elastic cross section 
in which the contribution of the continuum channels with \( I' \ne 0,2 \) 
has been 
excluded in an effective way. We recall that the elastic S-matrix in the USA 
is given by the sum of two terms, the first one coming from the folding 
potential and the latter, \( \Delta S \), which describes dynamic 
polarization effects due to the coupling to breakup channels. 
Therefore, this second term arises from the effect of the tidal forces.
However, the tidal potential \( v_T \) has vanishing diagonal matrix elements 
on the ground state of the projectile. Thus, 
the contribution of break-up states with angular momenta
$I'$ to the elastic matrix elements of \( \Delta S \) 
(up to lowest order in the tidal forces) depends on second order 
coupling through the expression
\begin{equation}
\langle \phi_{0}LIJ|\Delta S(I')|\phi_{0}LIJ\rangle \propto \sum_{bu,L'} \langle 
\phi_{0}LIJ|v_T|bu; L'I'J\rangle G^+(bu; L'I'J)
\langle bu; L'I'J|v_T|\phi_{0}LIJ\rangle ,
\label{eq39}
\end{equation}
where  \( G^+(bu; L' I' J) \) is a propagator and the sum extends to the 
breakup states.
In the particular case in which off-shell dependence is removed from the
the breakup states involved in the propagator, the right hand side of 
eq.~(\ref{eq39}) is proportional to the square of the distorted 
wave integral of the tidal potential. To lowest order, this term is
simply proportional to the break-up cross section.

A restriction in the number of breakup states considered will produce a 
reduction in the elastic matrix elements of \( \Delta S \). The results
presented in Fig.~2 show that the USA calculation including
$I'=0,2$ break-up is correct, whereas the calculation of break-up to larger
angular momentum is overestimated. Thus, making use of the
proportionality between elastic matrix elements of $\Delta S$ and the break-up
cross section, we get
\begin{equation}
\langle nLIJ|\Delta S(I'=0,2)|nLIJ\rangle =\langle nLIJ|\Delta S|nLIJ\rangle \frac{\sigma _{J}^{bu}(I'=0,2)}{\sigma _{J}^{bu}},
\end{equation}
where \( \sigma _{J}^{bu}(I'=0,2) \) is the restricted breakup cross section
and \( \Delta S(I'=0,2) \) its associated elastic S-matrix contribution. 
The corresponding elastic differential cross section 
resulting from this prescription, which we call ``restricted'' USA calculation,
is represented by the thick solid line in
Fig.\ \ref{fig:dni_e80_elxsec}. The result
is in very good agreement with the experimental data, supporting the hypothesis
that the strong coupling to high spin breakup states within USA
is responsible for the reduction in the elastic cross sections. If this 
coupling to high spin states $(I'>2)$ is excluded from the calculation, and the
elastic matrix elements are modified accordingly, then both the elastic 
differential cross sections and the partial break-up are well described.

\begin{figure}
{\par\centering \resizebox*{0.6\textwidth}{!}{\includegraphics{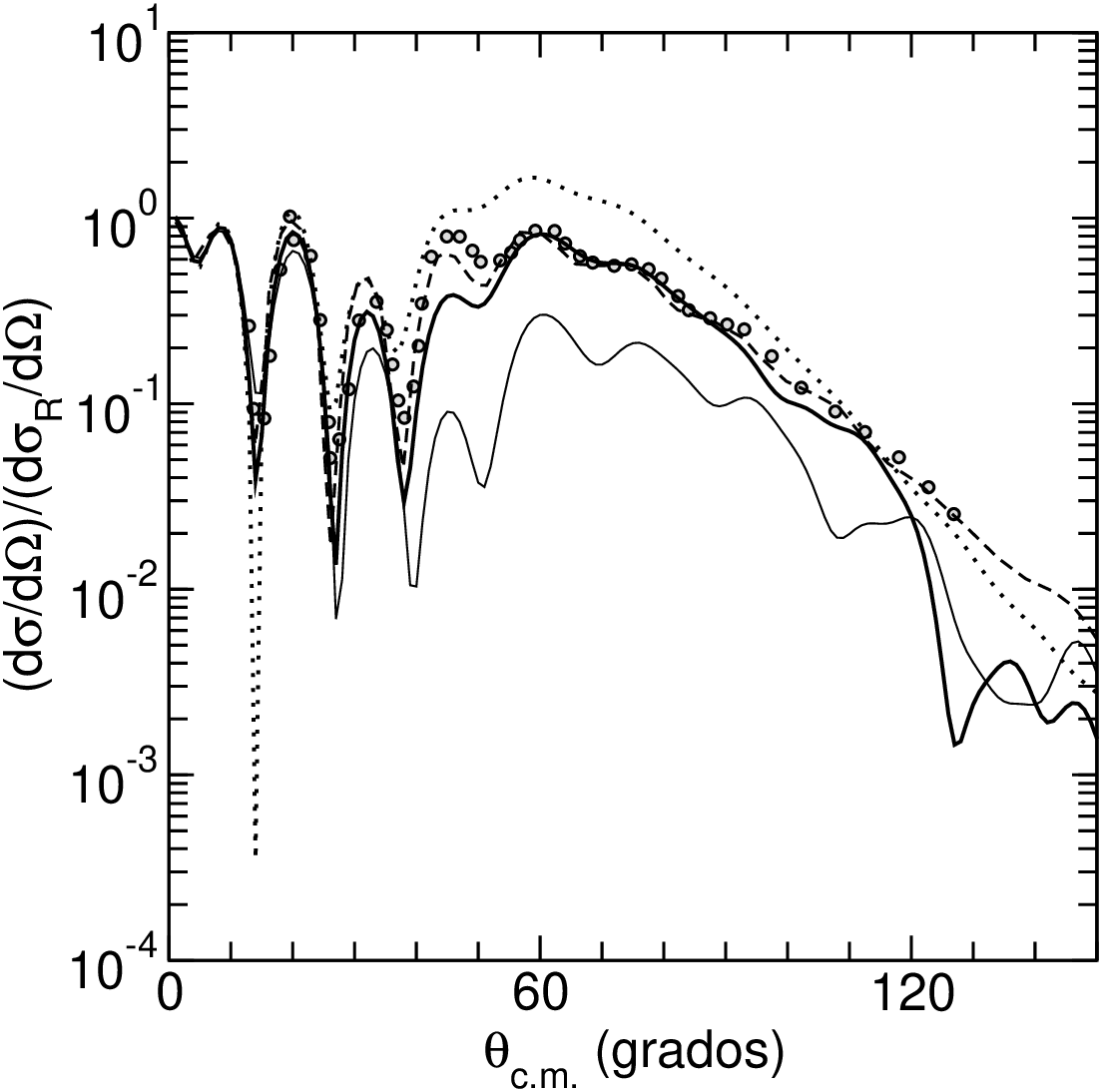}} \par}

\caption{\label{fig:dni_e56_elxsec}Elastic differential cross sections angular distributions
for \protect\( d\protect \) + \protect\( ^{58}\protect \)Ni at 56 MeV. The
meaning of the curves is the same as in Fig. \ref{fig:dni_e80_elxsec}.}
\end{figure}

\begin{figure}
{\par\centering \resizebox*{0.6\textwidth}{!}{\includegraphics{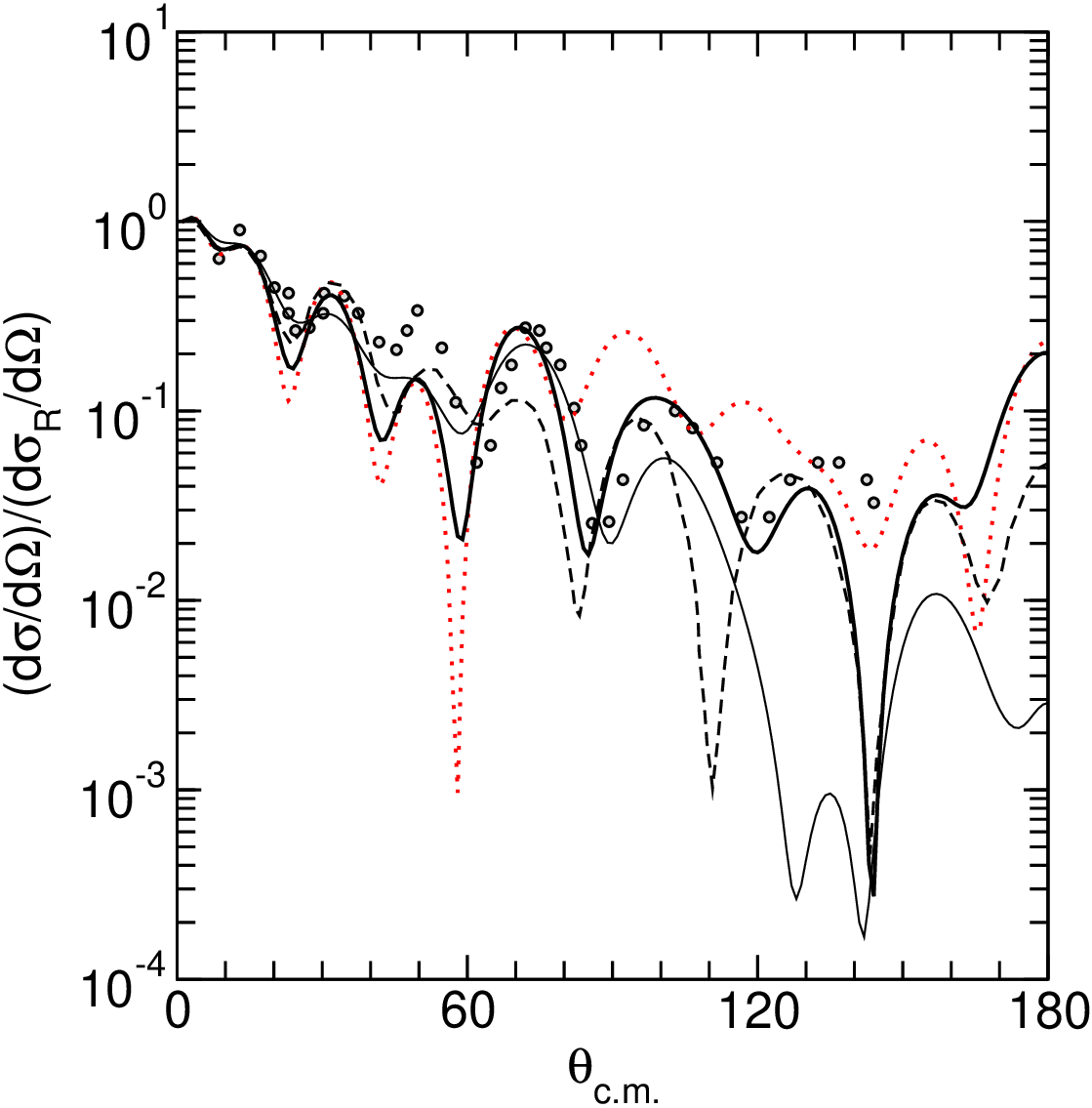}} \par}

\caption{\label{fig:dni_e22_elxsec}Elastic differential cross sections angular distributions
for \protect\( d\protect \) + \protect\( ^{58}\protect \)Ni at 21.6 MeV. The
meaning of the curves is the same as in Fig. \ref{fig:dni_e80_elxsec}.}
\end{figure}

An interesting question is to assess the validity of the USA at low energies
where other models, which are successfully applied to the high energy regime,
fail to reproduce the experimental data. This is the case of 
the sudden approximation and, in particular the Glauber model,
that can be regarded as a high energy approximation to the adiabatic
treatment. To this end we have applied the USA to the 
reaction \( d \) + \( ^{58} \)Ni
at 21.6 and 56 MeV, for which experimental data are available \cite{Per73}. 
We use
the same wave function for the deuteron ground state as in the 
previous calculations. The Perey optical model parameterization~\cite{Per63} 
for protons and neutrons has been selected.
The angular distribution of the elastic differential cross section
is shown in Figs.\ \ref{fig:dni_e56_elxsec} and \ref{fig:dni_e22_elxsec} 
where we compare the experimental data with the USA result (thin solid line), the
USA result considering break-up with \( I' =0, 2\) (thick solid line),  
the CDCC calculation (dashed line) and the folding model (dotted line). 

The results obtained for 56 MeV, shown in Fig.\ \ref{fig:dni_e56_elxsec},
are qualitatively similar to the ones for 80 MeV (Fig.~1). The
full USA calculation displays a too large
reduction in the elastic cross section with 
respect to the folding model prediction. However, 
the ``restricted'' USA calculation, which considers 
only the effect of break-up 
to \( I' =0, 2\), reproduces accurately the experimental data.

In Fig.\ \ref{fig:dni_e22_elxsec} we present the results for
21.6 MeV. Here there is a fairly good 
agreement between the full USA calculation
and the experiment up to 100 degrees. It is noticeable that the angular region
between 50 and 100 degrees is even better described by the USA model than by
the CDCC approach. The ``restricted'' USA calculation
improves the agreement with the experimental data. 

These results indicate that the USA is adequate to calculate deuteron 
break-up cross sections to states
with \( I'=0, 2 \), which are the most important break-up components, although
it overestimates the break-up cross sections to deuteron states with
\( I' > 2 \). As far as elastic scattering is concerned, the full USA
calculation gives in general, a too strong reduction in the cross sections.
However, the ``restricted'' USA calculation, which takes effectively 
into account
only the effect of coupling to \( I'=0, 2 \) break-up states, reproduces
satisfactorily deuteron elastic scattering data at all the energies 
considered.

Although more detailed tests should be required in order to
delimit the range of validity of the USA, these 
preliminary results suggest that
the model can be used as an alternative tool to analyze experimental data
of halo nuclei at relatively low energies, around and above the Coulomb 
barrier.

\section{Summary and conclusions}
 
We have developed a new approach, called the Uncorrelated Scattering 
Approximation (USA), applicable to the 
scattering of weakly bound nuclei. Our first requirement is that the target is
very heavy compared with the fragments of the projectile. Although 
this assumption is not essential to the model it simplifies
importantly the general formalism.
The USA arises from the fact that the three-body 
Hamiltonian corresponding to a composite projectile
can be written as a sum of two non-interacting two-body
Hamiltonians in two limit cases: i) when the tidal forces are neglected, and
ii) when the forces between the fragments of the projectile are neglected. 
In the first
case, the projectile remains in its ground state, the interaction with the 
target is given by the folding potential and the S-matrix is given by the
solution of the corresponding two-body scattering problem $S_F(L,E)$.
In the second case, the particles scatter independently. This means that
if the initial
wave function is given by the product of incident waves of the particles,
the wave function in the final state is characterized by
a product of outgoing waves, multiplied by
the corresponding S-matrices, $S_A(L_A,E_A)S_B(L_B,E_B)$.
Thus, within the USA one neglects tidal forces in the scattering region 
corresponding to large separations, whereas
in the scattering region, corresponding to small separations,
the forces between the fragments of the projectile are ignored.
The regions of large and small separations are defined in terms of the hyper
angular momentum $K$. Large separations correspond to $K>K_m$, and
small separations to $K\le K_m$, with $K_m$ defined such that its
turning point corresponds to a distance $R_m$ for which tidal forces
and the forces between the fragments are comparable. For $K>K_m$ the forces
between the fragments dominate, while for $K \le K_m$ tidal forces are
more important. Thus, $K_m$ is a parameter of the USA model. The other 
parameter, $\bar v$, is a constant that substitutes the interaction between the
fragments when $K<K_m$.

The proper boundary conditions for the region $K\le K_m$ requires
the projection of the initial state
in a discrete basis of states which are 
strongly localized for certain values of the relative
momentum of the fragments. These momentum localized states can be 
transformed analytically to states which have definite values of 
the hyper-angular momentum $K$.
Then, the Raynal-Revai coefficients are used
to transform the initial states, given in terms of the projectile-target
angular momentum $L$ and the internal angular momentum $I$, to states 
that depend directly
on the angular momentum of the fragments with the target, $L_A$ and $L_B$.
Finally, a certain combination of $K$-values is done 
in order to obtain states with
strongly localized values of the kinetic energy of each fragment. This
procedure
allows to write the 
S-matrix of the projectile in terms of the product 
$S_A(L_A,E_A)S_B(L_B,E_B)$ of the S-matrices of the fragments, evaluated at 
certain values of the energy and angular momentum of the fragments with 
respect to the target.
Then, the formalism can be applied directly
to describe scattering to bound and resonant
break-up states, as well as to direct break-up states.

The main advantage of the USA is that it allows to express
the S-matrix of the composite system in terms of the wave function of the
bound states and the scattering S-matrices of the fragments with the target.
These S-matrices can be easily obtained from calculations that fit the cross
sections of scattering experiments of the fragments with the target. In this 
sense, the fragment-target interaction is treated to all orders, and the
only assumption one makes on the reaction dynamics is to
neglect the correlation between the fragments. 
Thus, the USA approach presents some advantages compared with previous 
treatments.
Effects of absorption or ``shadowing'', which are included by means of
$ad-hoc$ profile functions in some spectator models \cite{GFJ96, GFJ97}, appear
naturally here, as occurs in Glauber approaches \cite{Brooke,Barranco}, 
because the S-matrices $S_A(L_A,E_A)$ and $S_B(L_B,E_B)$ 
of the fragments include the effects of absorption for low $L_A$ and $L_B$ 
values. Compared with semi-classical approaches \cite{Bertulani,Romanelli},
the USA does not make use of the concept of 
classical trajectories for
the relative motion. Compared with Glauber analyses \cite{Brooke,Barranco},
the USA  justifies the expression of the three-body 
S-matrix in terms of the product of the S-matrices of the fragments from
very general considerations, that do not require to make use of the eikonal
assumption of forward scattering, or straight line trajectories. Also, our
expressions depend on the S-matrices of the fragments evaluated at integer
orbital angular momenta, while previous Glauber-type approaches rely on 
evaluating by interpolation the S-matrices of the fragments as a function of 
real impact parameters, which would correspond to non-integer L-values.
Finally, the USA also differs from Glauber-type approaches, and 
other approaches based on the adiabatic approximation, for which the velocity
of all the fragments with respect to the target is equal to the projectile
velocity, and so the energies of the fragments are fixed quantities 
proportional to the masses. In the USA, the relative energy of the 
fragments with respect
to the target, $E_A^\ell$, $E_B^\ell$ can take different values,
reflecting the fact that
in the incident projectile the fragments have a certain momentum distribution.
As the USA does not make use of the eikonal approximation, which
is essentially a forward-scattering approach, we consider that it can be 
useful when applied to scattering of weakly bound halo 
nuclei at low scattering energies, comparable to the Coulomb barrier.

Apart from the theoretical formulation of the USA, we have also presented
a preliminary application of the approach to the reaction $d$+$^{58}$Ni 
at 21.6 and 80 MeV. The parameter $K_m$ is set to a large value, and $\bar v$
is taken as zero. This corresponds to neglect completely the effect of
proton-neutron correlation in the scattering.
Concerning elastic scattering at 80 MeV, the USA gives rise
to a depletion of the cross sections at intermediate 
angles with respect to the folding model prediction. The experimental data
and the CDCC calculations also show this depletion, although not so large as 
in the USA model. The break-up cross sections at 80 MeV to $I'=0$ and $I'=2$ 
states in the USA approach is consistent with the CDCC calculation. 
However, break-up cross sections to $I'>2$ states, which are sizeable in the 
USA, are strongly suppressed in the CDCC calculations. When the USA
calculation of the elastic scattering is modified to exclude effectively the
coupling to $I'>2$ states, the experimental differential cross 
sections and CDCC calculations are very well reproduced. Using the same 
procedure for the elastic cross sections at 21.6 and 56 MeV, the 
experimental data are also accurately reproduced.

Our interpretation of these results for deuteron scattering is that the
proton-neutron correlations, which are neglected in the USA calculation, play
a significant role. These correlations show up in the evaluation of break-up
cross sections to $I'>2$ states. Within the USA model, the proton and neutron 
scatter independently from the target, and so, after the scattering they have
a certain probability to end up in a state with large relative angular 
momentum. However, in reality the proton and neutron remain correlated, and 
this correlation suppresses large $I'$ break-up components in the wave 
function.
Once these components are excluded, both elastic and break-up
cross sections are well reproduced by the USA calculation.

We expect that, for the scattering of halo nuclei at energies around the 
Coulomb barrier, the effect of the 
correlations between the fragments will be less important than in the case of 
deuteron. Thus, the uncorrelated USA calculation will be more
reliable and we may expect to see more
clearly the features of the USA calculations. These include an important 
decrease of the elastic cross section with respect to the folding model 
prediction, and sizeable contributions to break-up with large angular momentum
between the fragments.

\bigskip

{\bf Acknowledgements }  We acknowledge fruitful discussions with 
J. Raynal and R.C. Johnson. This work has been partially supported by the 
Spanish CICyT project PB98-1111. A.M.M. 
acknowledges a grant from the Fundaci\'on
C\'amara of the Universidad de Sevilla. We are grateful to E. Stephenson and J.
Al-Khalili for providing us with the experimental data for the reaction $d$ + $^{58}$Ni in tabulated form.

\setcounter{section}{0} \catcode`\@=11 \@addtoreset{equation}{section}
\def\ksection{\Alph{section}}
\setcounter{equation}{0} \def\theequation{A.\arabic{equation}}
\catcode`@=12
\section*{Appendix A}

\subsection*{Momentum Localized States associated to hyper-spherical 
harmonics (HH)}

In this appendix we show how states characterized by definite 
values of the linear momenta of the constituents can be built starting from a 
truncated basis of hyper-spherical harmonics.

First, let us introduce the HH states, \( |K(L_{A}L_{B})JM_{J}\rangle  \), 
characterized
by the hyper-angular momentum \( K \), the orbital angular momenta 
\( L_{A} \) and \( L_{B} \) of particles $A$ and $B$, respectively, and 
the total angular momentum \( J \) and its projection
\( M_{J} \). We define the angle \( \beta  \) that connects the linear
momenta of the particles with the hyper-momentum \( \cal P \) 
through the relations
\( P_{A}=(\sqrt{M_{A}/M})\cal P\cos \beta  \) and 
\( P_{B}=(\sqrt{M_{B}/M})\cal P\sin \beta  \).
The states \( |K(L_{A}L_{B})JM_{J}\rangle  \) can be 
expressed in terms of states with definite values of \( \beta  \) as

\begin{equation}
\label{|K(L_{A}L_{B})JM>}
|K(L_{A}L_{B})JM_{J}\rangle =\int ^{\pi /2}_{0}d\beta \, f^{L_{A}L_{B}}_{K}(\beta )|\beta (L_{A}L_{B})JM_{J}\rangle, 
\end{equation}
 where 
\begin{equation}
\label{eq:f(LA,LB,K)}
f^{L_{A}L_{B}}_{K}(\beta )=N^{L_{A}L_{B}}_{K}(\cos \beta )^{L_{A}+1}(\sin \beta )^{L_{B}+1}P^{(L_{B}+\frac{1}{2},L_{A}+\frac{1}{2})}_{n}(\cos 2\beta ),
\end{equation}
where \( P_{n}^{(a,b)} \) denotes a Jacobi polynomial of degree 
\( n= (K-L_{A}-L_{B})/2\) and $N^{L_{A}L_{B}}_{K}$ represents
the normalization constant 

\begin{equation}
N^{L_{A}L_{B}}_{K}  =  \left[ \frac{2\, n!(K+2)(n+L_{A}+L_{B}+1)!}{\Gamma (n+L_{A}+\frac{3}{2})\Gamma (n+L_{B}+\frac{3}{2})}\right] ^{\frac{1}{2}}\\.
\end{equation}

The function \( f^{L_{A}L_{B}}_{K}(\beta ) \) is normalized to unity in the
variable in the variable \( \beta  \): 

\begin{equation}
\int _{0}^{\pi /2}d\beta \left[ f^{L_{A}L_{B}}_{K}(\beta )\right] ^{2}=1.
\end{equation}

This property also guarantees that the HH are normalized to unity. 
The hyper-angular
momentum can take the values \( K=L_{A}+L_{B},L_{A}+L_{B}+2,... \) giving
rise to an infinite number of states in eq.~(\ref{|K(L_{A}L_{B})JM>}). One
of the basic
ingredients in the Uncorrelated Scattering Approximation developed in
Section III is the introduction of a maximum
hyper-angular momentum \( K_{m} \) which reduces the infinite set of states
to a finite one: \( K=L_{A}+L_{B},L_{A}+L_{B}+2,...,K_{m} \). In terms of the
index \( n \) we have the subset: \( n=0,1,...,N-1 \), with 
\( N=(K_{m}-L_{A}-L_{B})/2+1 \).

The family of Jacobi polynomials, \{\( P^{(a,b)}_{n}(x);n=0,...,N-1 \)\}, 
constitute
an orthogonal set of functions in the interval (\( -1,+1) \) with respect to
the weight function 
\( \omega (x)=\frac{1}{4}(\frac{1-x}{2})^{a}(\frac{1+x}{2})^{b} \): 

\begin{equation}
\int ^{1}_{-1}dx\, \omega (x)[P^{(a,b)}_{n}(x)]^{2}=h_{n}
\end{equation}
 where

\begin{equation}
h_{n}=\frac{\Gamma [n+a+1]\Gamma [n+b+1]}{2(2n+a+b+1)n!\Gamma [n+a+b+1]} .
\end{equation}

From this family of Jacobi polynomials a new set of \( N \) polynomials of degree
\( N-1 \) are defined as~\cite{Per99} 

\begin{equation}
Q^{(a,b)}_{N-1}(x_{\ell },x)=\sum ^{K_{m}}_{K=L_{A}+L_{B}}\left[ N^{L_{A}L_{B}}_{K}\right] ^{2}P^{(a,b)}_{n}(x)P^{(a,b)}_{n}(x_{\ell });\, \, \, \ell =1,\ldots ,N.
\end{equation}

Here, \( \{x_{\ell };\ell =1,2,\ldots ,N\} \) represent the zeros of the polynomial
\( P_{N}^{(a,b)}(x). \) Direct application of Christoffel-Darboux formula 
(see ref.~\cite{Abra72}) on the previous expression leads to 

\begin{equation}
Q^{(a,b)}_{N-1}(x_{\ell },x)=\frac{P^{(a,b)}_{N-1}(x_{\ell })}{h_{N-1}}\frac{k_{N-1}}{k_{N}}\frac{P_{N}^{(a,b)}(x)}{(x-x_{\ell })}
\end{equation}
 where \( k_{N} \) is the coefficient of \( x^{N} \) in \( P_{N}^{(a,b)}(x) \).
The polynomials \( Q^{(a,b)}_{N-1}(x_{\ell },x) \) are orthogonal in the same
interval and relative to the same weight function as the original polynomials.
The expression above shows that the polynomial \( Q^{(a,b)}_{N-1}(x_{\ell },x) \)
vanishes at the points \( x=x_{s};s=1,\ldots ,N \) except at \( x=x_{\ell } \).

The new set of polynomials allows us to construct Momentum Localized States
(MLS) which are defined as

\begin{eqnarray}
|\ell (L_{A}L_{B})JM_{J}\rangle  & = & (w_{N\ell }^{L_{A}L_{B}})^{(1/2)}\int ^{\pi /2}_{0}d\beta (\cos \beta )^{L_{A}+1}(\sin \beta )^{L_{B}+1}\nonumber \\
 & \times  & Q^{(L_{B}+\frac{1}{2},L_{A}+\frac{1}{2})}_{N-1}(x _{\ell ,}\cos (2\beta) )|\beta (L_{A}L_{B})JM_{J}\rangle 
\end{eqnarray}
with

\begin{equation}
\label{eq:w_s(Q)}
w^{(L_{A}L_{B})}_{N\ell }=\left[ Q^{(L_{B}+\frac{1}{2},L_{A}+\frac{1}{2})}_{N-1}(x _{\ell },x _{\ell })\right] ^{-1}.
\end{equation}

Substituting the explicit expressions of \( h_{N-1} \), \( k_{N-1} \) 
and \( k_{N} \)
\cite{Abra72} in (\ref{eq:w_s(Q)}) one gets

\begin{equation}
w^{(L_{A}L_{B})}_{N\ell }=\frac{1}{2^3}\frac{(2N+a+b)^{2}}{(N+a)^{2}(N+b)^{2}}\frac{\Gamma[N+a+1]\Gamma[N+b+1]}{N!\,\, \Gamma[N+a+b+1]}\frac{1-x^{2}_{\ell }}{\left[ P^{(a,b)}_{N-1}(x_{\ell })\right] ^{2}},
\end{equation}
 where \( a=L_{B}+1/2 \) and \( b=L_{A}+1/2 \) in our case. 

The state \( |\ell (L_{A}L_{B})JM_{J}\rangle  \) is
characterized by values of the momenta of the two particles
sharply peaked around 
\( P^{\ell }_{A}=(\sqrt{M_{A}/M}){\cal P}\sqrt{(1+ x_{\ell })/2} \)
and \( P^{\ell }_{B}=(\sqrt{M_{B}/M}){\cal P}\sqrt{(1- x_{\ell })/2} \). 
This localization
is enhanced as the number of states increases which, in turn, depends on the
cut off hyper-angular momentum \( K_{m} \).

An orthogonal transformation can be defined between the truncated 
hyper-spherical basis and the MLS basis:

\begin{equation}
|\ell (L_{A}L_{B})JM_{J}\rangle =\sum ^{K_{m}}_{K=L_{A}+L_{B}}\langle K|\ell \rangle _{L_{A}L_{B}}|K(L_{A}L_{B})JM_{J}\rangle .
\end{equation}

This relation can be inverted allowing to express the HH in terms of 
the MLS:

\begin{equation}
|K(L_{A}L_{B})JM_{J}\rangle =\sum ^{N}_{\ell=1 }\langle \ell |K\rangle _{L_{A}L_{B}}|\ell (L_{A}L_{B})JM_{J}\rangle .
\end{equation}

The coefficients of the transformation can be computed from the definition of
the MLS and the polynomials \( Q_{N-1}(x_{\ell},\cos (2\beta) ) \):

\begin{equation}
\langle K|\ell \rangle _{L_{A}L_{B}}=(w_{N\ell }^{L_{A}L_{B}})^{1/2}N^{L_{A}L_{B}}_{K}P^{(L_{B}+1/2,L_{A}+1/2)}_{n}(x_{\ell }).
\end{equation}

\begin{figure}
{\par\centering \resizebox*{0.7\textwidth}{!}{\includegraphics{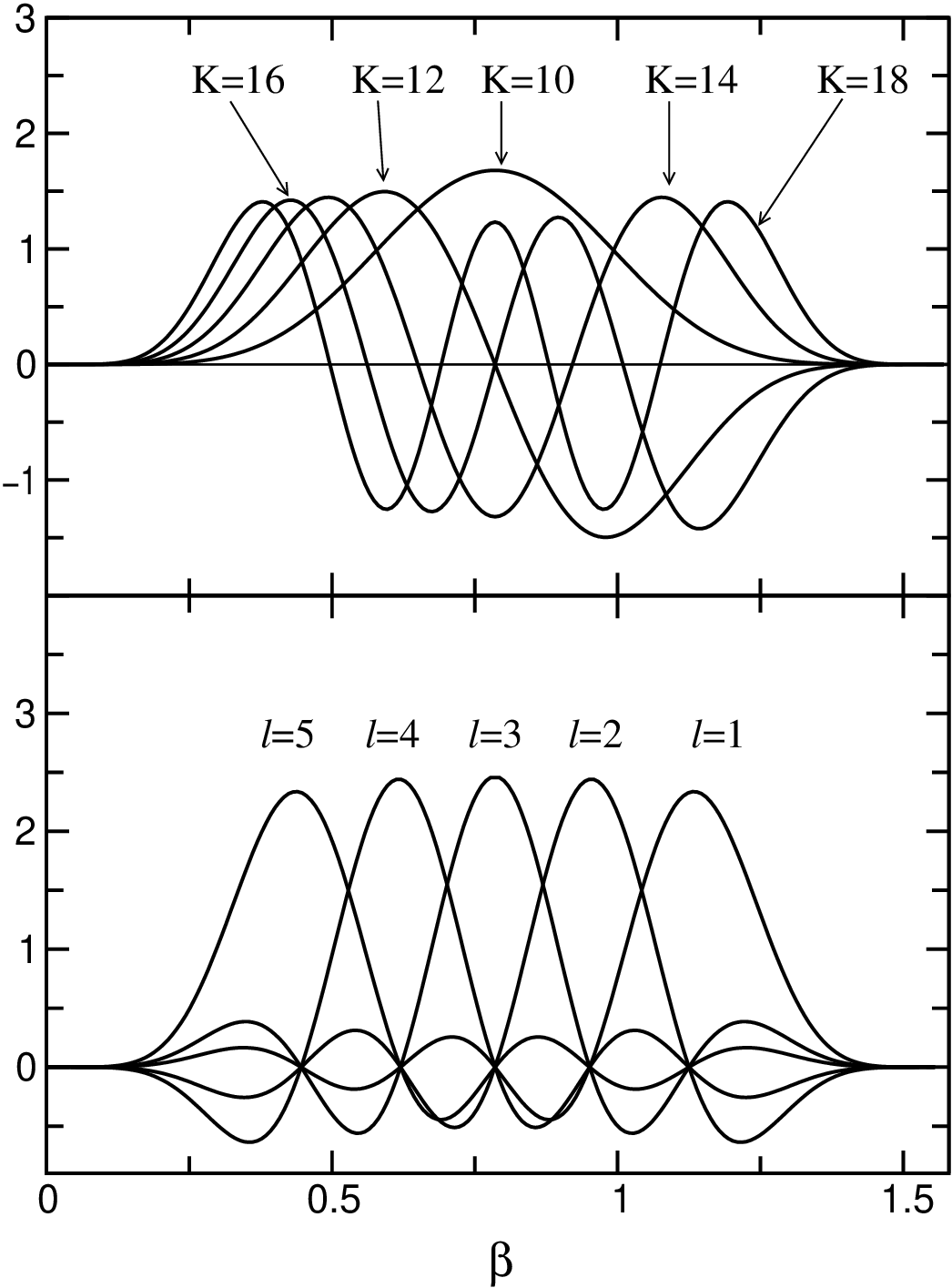}} \par}

\caption{\label{Fig:cls} $f_{K}^{L_{A}L_{B}}(\beta)$ versus the variable $\beta$ 
for $L_{A}=L_{B}$=5 and $K$=10,12,14,16 and 18 (upper figure) and
associated Momentum Localized States, labeled by the index $\ell$, defined in the text (bottom figure). }
\end{figure}

In Fig.~\ref{Fig:cls} the functions 
$f_{K}^{L_{A}L_{B}}(\beta)$  (upper part) and the corresponding
MLS (lower part) are plotted versus the variable $\beta$  for the case
$L_{A}$=$L_{B}$=5. A basis with $N$=5 states has been used,
corresponding to the values of the hyper-angular momentum 
$K$=10 to 18, in units of two. The MLS are labeled
with the index $\ell$ which is associated with the roots
of the Jacobi polynomial $P_{5}^{(5+\frac{1}{2},5+\frac{1}{2})}(x)$. Thus, 
the MLS $\ell=1$ is localized around $x=-0.64$, that corresponds to 
$\beta=1.13$. In terms of the energy of the two particles this means
that the fraction of the total available kinetic energy carried by 
particles $A$ and $B$ are $E_{A}/(E-\bar{v})=0.18$ and 
$E_{B}/(E-\bar{v})=0.82$, respectively. This is in fact 
the most asymmetric situation
for this value of $K_{m}$. On other side, the MLS $\ell=3$ corresponds to a 
physical situation in which both particles carries half of the total
kinetic energy.

In a similar way, it is possible to define HH with definite values of 
the hyper-angular
momentum \( K \), orbital momentum \( L \), intrinsic spin \( I \)
and total angular momentum \( J \) and  projection \( M_{J} \). These new
HH, denoted by \( |K(LI)JM_{J}\rangle  \), can be related to the set \( |K(L_{A}L_{A})JM_{J}\rangle  \)
by means of the Raynal-Revai transformation \cite{Raynal}:

\begin{equation}
|K(LI)JM_{J}\rangle =\sum _{L_{A}L_{B}}\langle L_{A}L_{B}|LI\rangle _{JK}|K(L_{A}L_{B})JM_{J}\rangle. 
\end{equation}

Introducing the angle \( \alpha  \), which denotes the ratio between the
internal momentum \( p \) and the hyper-momentum 
\( \cal P \), \( p={\cal P} \sqrt{m/M}\sin \alpha \),
it is possible to expand the new HH as

\begin{equation}
\label{|K(LI)JM>}
|K(LI)JM_{J}\rangle =\int ^{\pi /2}_{0}d\alpha \, f^{LI}_{K}(\alpha )
|\alpha (LI)JM_{J}\rangle, 
\end{equation}
where the function \( f^{LI}_{K}(\alpha ) \) is given by the analogous to
(\ref{eq:f(LA,LB,K)}), i.e.

\begin{equation}
\label{eq:f(I,L,K)}
f^{LI}_{K}(\alpha )=N^{LI}_{K}(\cos \alpha )^{L+1}
(\sin \alpha )^{I+1}P^{(I+\frac{1}{2},L+\frac{1}{2})}_{n}(\cos 2\alpha ),
\end{equation}
with $n=(K-L-I)/2$ in this case.

Introducing a cut-ff hyper-angular momentum,  $K_m$, it is possible to 
construct a set of Momentum Localized States associated with the truncated
basis \( |K(LI)JM_{J}\rangle  \), $L+I<K<K_{m}$. They are developed in 
terms of the angle $\alpha$ as

\begin{eqnarray}
|j(LI)JM_{J}\rangle  & = & (w_{Nj}^{LI})^{(1/2)}\int ^{\pi /2}_{0}d\alpha (\cos \alpha )^{L+1}(\sin \alpha )^{I+1}\nonumber \\
 & \times  & Q^{(I+\frac{1}{2},L+\frac{1}{2})}_{N-1}(x_{j},\cos 2\alpha )|\alpha (LI)JM_{J}\rangle 
\end{eqnarray}
where  \( N \) is the number of values of \( K \) of the HH truncated basis 
and \( \{x_{j};j=1,\ldots ,N\} \) represent the zeros of the polynomial
\( P_{N}^{(I+\frac{1}{2},L+\frac{1}{2})}(x) \). The state 
\( |j(LI)JM_{J}\rangle  \)
is characterized by narrow distributions of the relative and center of mass
momenta around the central values 
\( p_{j}=\sqrt{\frac{\mu }{M}}\mathcal{P}\sqrt{(1-x_{j})/2} \)
and \( P_{j}=\mathcal{P}\sqrt{(1+x_{j})/2} \).

The original basis and the MLS basis are related by means of the orthogonal
transformation

\begin{equation}
|K(LI)JM_{J}\rangle =\sum ^{N}_{j=1}\langle j|K\rangle _{LI}|j(LI)JM_{J}\rangle ,
\end{equation}
where the transformation coefficients are given by 
\begin{equation}
\langle K|j\rangle _{LI}=(w_{Nj}^{LI})^{1/2}N^{LI}_{K}P^{(I+1/2,L+1/2)}_{n}(x_{j}).
\end{equation}


\end{document}